*M. Altoumaimi,*

*V.V. Loboda, Dr.Sci (Phys.-Math.)*

# A RIGID BEAM ACTING IN THE SHEARING MANNER TO THE QUASI-CRYSTALLINE HALF-SPACE

**An analytical model describing the action a rigid beam under shear forces to the quasi-crystalline half-space is analyzed. The elasticity theory of quasicrystals and the method of complex variables is used. Graphs of stresses and displacements for both phonon and phason fields along the edge of the half-space have been obtained.**

**Introduction.** Quasicrystals, found by Shechtman et al. [6] are a new class of materials which are actively studied for about several last decades. Quasicrystals (QCs) differ from ordinary crystals and non-crystals by their high strength, high wear resistance, low heat-transfer, etc. These materials, are nowadays used in in coating surface of engines, solar cells, thermoelectric converters, containers of nuclear fuel, etc.

Because of the quasi-periodic symmetry of QCs, concepts of the high-dimensional space have been introduced instead of the classical crystallographic theory. The phonon field represents the lattice vibrations in QCs, and the phason field defines the quasi-periodic rearrangement of atoms. Both these fields are used to describe the elastic properties of QCs. Theory of elasticity of QCs and the state of the art of this subject is presented in [1-3].

One-dimensional (1D) QCs exhibit just one quasi-periodic axis, while the perpendicular plane reveals classical crystalline properties. The apace group of such QCs is analyzed in [8] and the problems of cracks were investigated in the references [7, 9]. Frictional contact of one-dimensional hexagonal quasicrystal coating with an account of thermal effects was investigated in [84]. The action of the rigid punch in the shearing way to the semi-infinite quasicrystal space has never been considered earlier to the author's knowledge. Such problem is considered in the present paper.

**Formulation of the basic relations.** For the linear elastic theory of QCs, the constitutive relations have the following form [4]

$$\sigma_{ij} = c_{ijks}\varepsilon_{ks} + R_{ij3s}w_{3s}, \tag{1}$$

$$H_{3i} = R_{ks3i}\varepsilon_{ks} + K_{3i3s}w_{3s}, \tag{2}$$

where $i, j, k, s = 1, 2, 3$, and the denotation "," represents the derivative operation for the space variables; $u_i$, $w_3$ are the phonon displacements, and phason displacement, respectively, and the atom arrangement is periodic in the $x_1 - x_2$ plane and quasi-periodic in the $x_3$-axis; $\sigma_{ij}$ and $\varepsilon_{ks}$ are the phonon stresses and strains, respectively; $H_{3i}$ and $w_{3i}$ are the phason stresses and strains, respectively; $c_{ijks}$ and $K_{3j3s}$ are the elastic constants in the phonon and phason fields, respectively; $R_{ij3k}$ represent the phonon–phason coupling elastic constants; Here, a comma in subscript denotes differentiation with respect to the following spatial variable.

The phonon displacements $u_1$, $u_2$, $u_3$ and phason displacement $w_3$ ($w_1 = w_2 = 0$) of 1D hexagonal QCs are connected with corresponding deformations in the following way

$$\varepsilon_{11} = \frac{\partial u_1}{\partial x_1}, \quad \varepsilon_{22} = \frac{\partial u_2}{\partial x_2}, \quad \varepsilon_{33} = \frac{\partial u_3}{\partial x_3}, \tag{3}$$

$$\varepsilon_{23} = \varepsilon_{32} = \frac{1}{2}\left(\frac{\partial u_3}{\partial x_2} + \frac{\partial u_2}{\partial x_3}\right), \quad \varepsilon_{31} = \varepsilon_{13} = \frac{1}{2}\left(\frac{\partial u_3}{\partial x_1} + \frac{\partial u_1}{\partial x_3}\right),$$

$$\varepsilon_{12} = \varepsilon_{21} = \frac{1}{2}\left(\frac{\partial u_1}{\partial x_2} + \frac{\partial u_2}{\partial x_1}\right), \quad w_{13} = \frac{\partial w_3}{\partial x_1}, \quad w_{32} = \frac{\partial w_3}{\partial x_2}, \quad w_{33} = \frac{\partial w_3}{\partial x_3} \tag{4}$$

whilst other $w_{ij} = 0$.

Further we'll use the contracted notations for the presentation of the relations (1), (2), in which the following exchange of indices takes place $11 \to 1$, $22 \to 2$, $33 \to 3$, $23 \to 4$, $31 \to 5$, $12 \to 6$ and $C_{ijkl}$ is designated as $C_{pq}$. Thus taking into account that 1D hexagonal QCs have transversally-isotropic structure we can write

$$C_{11} = C_{1111} = C_{2222}, \quad C_{12} = C_{1122}, \quad C_{33} = C_{3333}, \quad C_{44} = C_{2323} = C_{3131},$$

$$C_{13} = C_{1133} = C_{2233}, \quad C_{66} = \frac{C_{11} - C_{12}}{2} = \frac{C_{1111} - C_{1122}}{2}.$$

Thus, there are five independent phonon elastic constants. In addition, there are two independent phason elastic constants $K_1 = K_{3333}$, $K_2 = K_{3131} = K_{3232}$, as well as three elastic coupling constants of phonon and phason fields $R_1 = R_{1133} = R_{2233}$, $R_2 = R_{3333}$, $R_3 = R_{2332} = R_{3131}$.

In the new notations the constitutive relations between stresses and strains can be written in the form [2]:

$$\begin{cases} \sigma_{11} = C_{11}\varepsilon_{11} + C_{12}\varepsilon_{22} + C_{13}\varepsilon_{33} + R_1 w_{33}, \\ \sigma_{22} = C_{12}\varepsilon_{11} + C_{11}\varepsilon_{22} + C_{13}\varepsilon_{33} + R_1 w_{33}, \\ \sigma_{33} = C_{13}\varepsilon_{11} + C_{13}\varepsilon_{22} + C_{33}\varepsilon_{33} + R_2 w_{33}, \\ \sigma_{23} = \sigma_{32} = 2C_{44}\varepsilon_{23} + R_3 w_{32}, \\ \sigma_{31} = \sigma_{13} = 2C_{44}\varepsilon_{31} + R_3 w_{31}, \\ \sigma_{12} = \sigma_{21} = 2C_{66}\varepsilon_{12}, \\ H_{33} = R_1\left(\varepsilon_{11} + \varepsilon_{22}\right) + R_2\varepsilon_{33} + K_1 w_{33}, \\ H_{31} = 2R_3\varepsilon_{31} + K_2 w_{31}, \\ H_{32} = 2R_3\varepsilon_{23} + K_2 w_{32}, \end{cases} \quad (5)$$

and other $H_{ij} = 0$.

The equilibrium equations are as follows:

$$\frac{\partial \sigma_{11}}{\partial x_1} + \frac{\partial \sigma_{12}}{\partial x_2} + \frac{\partial \sigma_{13}}{\partial x_3} = 0, \quad \frac{\partial \sigma_{21}}{\partial x_1} + \frac{\partial \sigma_{22}}{\partial x_2} + \frac{\partial \sigma_{23}}{\partial x_3} = 0, \quad \frac{\partial \sigma_{31}}{\partial x_1} + \frac{\partial \sigma_{32}}{\partial x_2} + \frac{\partial \sigma_{33}}{\partial x_3} = 0,$$

$$\frac{\partial H_{31}}{\partial x_1} + \frac{\partial H_{32}}{\partial x_2} + \frac{\partial H_{33}}{\partial x_3} = 0, \quad (6)$$

For the case of anti-plane mechanical loading with reference to the $x_1 O x_2$-plane all fields are independent of the variable $x_3$. Therefore, the problem under consideration is a so-called anti-plane shear problem. In this case

$$u_1 = u_2 = 0, \quad u_3 = u_3(x_1, x_2), \quad w_3 = w_3(x_1, x_2)$$

and the constitutive relations take the form:

$$\begin{Bmatrix} \sigma_{j3} \\ H_{j3} \end{Bmatrix} = \mathbf{R} \begin{Bmatrix} u_{3,j} \\ w_{3,j} \end{Bmatrix} \quad (j = 1, 2), \quad (7)$$

where $\mathbf{R} = \begin{bmatrix} c_{44} & R_3 \\ R_3 & K_2 \end{bmatrix}$, (8)

and $c_{44}, K_2, R_3$ stand for the phonon elastic modulus, phason elastic modulus and phonon-phason coupling modulus, respectively; which are written in the simplified index notation. Introducing the designations

$$\mathbf{u} = [u_3, w_3]^T, \quad \mathbf{t}_j = [\sigma_{3j}, H_{3j}]^T, \quad (9)$$

one can write

$$\mathbf{t}_j = \mathbf{R} \mathbf{u}_{,j} \quad (j = 1, 2). \quad (10)$$

For the considered anti-plane problem the equilibrium equations (4) takes the form

$$\frac{\partial \sigma_{31}}{\partial x_1} + \frac{\partial \sigma_{32}}{\partial x_2} = 0, \quad \frac{\partial H_{31}}{\partial x_1} + \frac{\partial H_{32}}{\partial x_2} = 0.$$

Substituting (7) in the last equation we get that the functions $u_3$ and $w_3$ satisfy the equations $\Delta u_3 = 0$, $\Delta w_3 = 0$, respectively, i.e. they are harmonic. Therefore, present the vector $u$, composed of these function, as real parts of some analytic vector-function

$$u = 2\operatorname{Re}\Phi(z) = \Phi(z) + \bar{\Phi}(\bar{z}), \qquad (11)$$

where $\Phi(z) = [\Phi_1(z), \Phi_2(z)]^T$ is an arbitrary analytic vector-function of the complex variable $z = x_1 + ix_2$.

Substituting (11) in (10), one gets

$$t_1 = -iB\Phi'(z) + i\bar{B}\bar{\Phi}'(\bar{z}), \quad t_2 = B\Phi'(z) + \bar{B}\bar{\Phi}'(\bar{z}), \qquad (12)$$

where $B = iR$.

**A stamp acting in a shearing way to the quasicrystaline half-space.** Consider a beam acting to the quasicrystaline half-space (Fig. 1) under the action of the shear loading $\mathbf{P} = [P_1, P_2]^T$, which value does not change along the axis $x_3$. First component $P_1$ correspond to the phonon loading while second component $P_2$ - to the phason loading. In this case the out of plane state takes place and only the cross-section in the plane $(x_1, x_2)$ can be considered (Fig. 2).

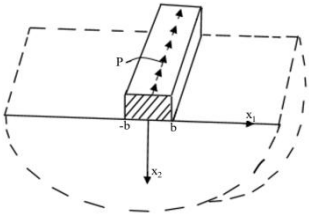 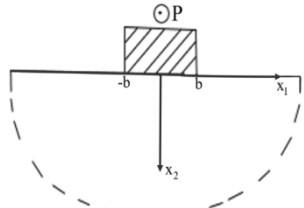

Fig. 1 – **Rigid beam acting to the quasicrystaline half-space**

Fig. 2 – **Cross-section of composition for out of plane problem formulation**

The formulas (7) and (8) hold true in this case and the presentations (11) and (12) can be used. The boundary conditions for the considered problem can be written in the form

$$\mathbf{t}_2(x_1, 0) \equiv \begin{Bmatrix} \sigma_{23}(x_1, 0) \\ H_{23}(x_1, 0) \end{Bmatrix} = \mathbf{0} \text{ for } /x_1| > b \qquad (13)$$

$$\mathbf{u}_{,1}(x_1, 0) \equiv \left[\frac{\partial u_3(x_1, 0)}{\partial x_1}, \frac{\partial w_3(x_1, 0)}{\partial x_1}\right]^T = 0 \text{ for } /x_1| \leq b \qquad (14)$$

As was mentioned above the vector-function $\Phi(z)$ is analytic in the lower half-plane, i.e. For $x_2 < 0$. We denote the lower half-plane as $S^-$. Let's

redefine this vector-function to the upper half-plane, which we denote as $S^+$, on the following rule:

$$\mathbf{\Phi}'(z) = \bar{\mathbf{\Phi}}'(z), \text{ for } z \in S^+, \tag{15}$$

where according to [5]

$$\bar{\mathbf{\Phi}}'(z) = \overline{\mathbf{\Phi}'(\bar{z})}.$$

Let's write $\bar{z}$ instead of $z$ considering that $z \in S^-$

$$\mathbf{\Phi}'(\bar{z}) = \bar{\mathbf{\Phi}}'(\bar{z})$$

and take conjugation from the obtained relation. We arrive to the equation

$$\bar{\mathbf{\Phi}}'(z) = \mathbf{\Phi}'(z) \text{ for } z \in S^-.$$

Taking conjugation one more time, we arrive to the equation

$$\mathbf{\Phi}'(\bar{z}) = \overline{\mathbf{\Phi}'(z)} \text{ for } z \in S^- \tag{16}$$

Considering the derivative on $x_1$ of Eq. (11)

$$\boldsymbol{u}'_{,1} = \mathbf{\Phi}'(z) + \bar{\mathbf{\Phi}}'(\bar{z})$$

and using the formula (16), we get

$$\boldsymbol{u}_{,1} = \mathbf{\Phi}'(z) + \mathbf{\Phi}'(\bar{z}) \text{ for } z \in S^- \tag{17}$$

Due to (16) the second formula (12) can be presented as

$$\boldsymbol{t}_2 = \boldsymbol{B}\mathbf{\Phi}'(z) + \bar{\boldsymbol{B}}\mathbf{\Phi}'(\bar{z}) \text{ for } z \in S^- \tag{18}$$

Tending $z$ to $x_1 - i0$, i.e. to the boundary of the region from the down, the equations (17) and (18) attain the form

$$\boldsymbol{u}^-_{,1}(x_1, 0) = \mathbf{\Phi}'^-(x_1) + \mathbf{\Phi}'^+(x_1), \tag{19}$$

$$\boldsymbol{t}^-_2(x_1, 0) = \boldsymbol{B}\mathbf{\Phi}'^-(x_1) + \bar{\boldsymbol{B}}\mathbf{\Phi}'^+(x_1) \tag{20}$$

where $\mathbf{\Phi}'^{\pm}(x_1) = \lim\limits_{x_2 \to 0} \mathbf{\Phi}(x_1 \pm ix_2)$.

Next we take into account that $\boldsymbol{B} = i\boldsymbol{R}$ and $\bar{\boldsymbol{B}} = -i\boldsymbol{R}$ because the matrix $\boldsymbol{R}$ is real. Then from boundary condition (13) and Eq. (20) we can conclude that the function $\mathbf{\Phi}(z)$ is analytically continued via the sections $/x_1|>b$ of the half-plane edge. It means that this function is analytic in the whole complex plane except the segment $[-b, b]$.

**Formulation and solution of the problem of linear relationship.** Satisfying the boundary condition (14) with use of (19) we arrive to the following equation

$$\mathbf{\Phi}'^-(x_1) + \mathbf{\Phi}'^+(x_1) = 0 \text{ for } /x_1|\leq b \tag{21}$$

Additional condition for the Eq. (21) follows from the equation of equilibrium

$$\int_{-b}^{b} \boldsymbol{t}^-_2(x_1, 0) dx = \boldsymbol{P} \tag{22}$$

Accounting that

$$\mathbf{\Phi}'^{+}(x_1) = -\mathbf{\Phi}'^{-}(x_1) \text{ for } /x_1/\leq b,$$

We get due to (20)

$$t_2^-(x_1,0) = (\mathbf{B}-\bar{\mathbf{B}})\mathbf{\Phi}'^{-}(x_1) \text{ for } /x_1/\leq b \qquad (23)$$

And equation (22) can be rewritten in the form

$$\int_{-b}^{b} \mathbf{\Phi}'^{-}(x_1)dx = (\mathbf{B}-\bar{\mathbf{B}})^{-1}\mathbf{P}. \qquad (24)$$

Due to [5] the solution of Eq. (21) disappearing at infinity can be presented in the form

$$\mathbf{\Phi}'(z) = \frac{\mathbf{C}_0}{\sqrt{z^2-b^2}}$$

Arbitrary vector-constant $\mathbf{C}_0$ can be found from (24) by the following calculations:

$$\mathbf{C}_0 \int_{-b}^{b} \frac{1}{i\sqrt{b^2-x^2}}dx = (\mathbf{B}-\bar{\mathbf{B}})^{-1}\mathbf{P},$$

$$\mathbf{C}_0 \frac{\pi}{i} = (\mathbf{B}-\bar{\mathbf{B}})^{-1}\mathbf{P}, \quad \mathbf{C}_0 = \frac{(\mathbf{B}-\bar{\mathbf{B}})^{-1}\mathbf{P}}{\pi}i.$$

It gives the following expression for $\mathbf{\Phi}'(z)$:

$$\mathbf{\Phi}'(z) = i\frac{(\mathbf{B}-\bar{\mathbf{B}})^{-1}\mathbf{P}}{\pi\sqrt{z^2-b^2}}.$$

Taking into account that $\mathbf{B} = i\mathbf{R}$ and the matrix $\mathbf{R}$ is real, we have $\mathbf{B}-\bar{\mathbf{B}} = 2\mathbf{B} = 2i\mathbf{R}$. It means that $(\mathbf{B}-\bar{\mathbf{B}})^{-1} = -0.5i\mathbf{R}^{-1}$ and therefore

$$\mathbf{\Phi}'(z) = \frac{\mathbf{R}^{-1}\mathbf{P}}{2\pi\sqrt{z^2-b^2}} \qquad (25)$$

Integrating the last formula, we obtain:

$$\mathbf{\Phi}(z) = \frac{\mathbf{R}^{-1}\mathbf{P}}{2\pi}\left\{\frac{1}{2}\left[\log\left(\frac{z}{\sqrt{z^2-b^2}}+1\right) - \log\left(1-\frac{z}{\sqrt{z^2-b^2}}\right)\right] + \frac{C^*}{2}\right\}, \qquad (26)$$

where $C^*$ is an arbitrary constant.

**Determination of the phonon and phason characteristics at the boundary.** Substituting (25) in (19) and taking into account that

$$\sqrt{z^2-b^2} = \sqrt{x_1^2-b^2} \text{ for } z \to x_1^- \text{ for } x_1 > b$$

and $\mathbf{\Phi}'^{-}(x_1) = \mathbf{\Phi}'^{+}(x_1)$ for $/x_1/>b$, one gets

$$u_{,1}(x_1,0) = 2\mathbf{\Phi}'^{-}(x_1) = \frac{\mathbf{R}^{-1}\mathbf{P}}{\pi\sqrt{x_1^2 - b^2}} \text{ for } x_1 > b. \quad (27)$$

Further from the formulas (11) and (26) we can write

$$\mathbf{u} = \frac{\mathbf{R}^{-1}\mathbf{P}}{\pi}\text{Re}\left\{\left[\log\left(\frac{z}{\sqrt{z^2-b^2}}+1\right) - \log\left(1 - \frac{z}{\sqrt{z^2-b^2}}\right)\right] + C^*\right\}. \quad (28)$$

Substituting formula (25) in (23) one gets

$$\mathbf{t}_2^{-}(x_1,0) = i\frac{\mathbf{P}}{\pi\sqrt{x_1^2 - b^2}} = \frac{\mathbf{P}}{\pi\sqrt{b^2 - x_1^2}} \text{ for } /x_1/\leq b.$$

It means that

$$\sigma_{32}(x_1,0) = \frac{P_1}{\pi\sqrt{b^2 - x_1^2}}, \ H_{32}(x_1,0) = \frac{P_2}{\pi\sqrt{b^2 - x_1^2}} \text{ for } /x_1/\leq b. \quad (29)$$

Taking into account that according to (8)

$$\mathbf{R}^{-1}\mathbf{P}_0 = \begin{bmatrix} c_{44} & R_3 \\ R_3 & K_2 \end{bmatrix}^{-1}\mathbf{P}_0 = \frac{1}{\Delta}\begin{bmatrix} K_2 & -R_3 \\ -R_3 & c_{44} \end{bmatrix}\begin{bmatrix} P_1 \\ P_2 \end{bmatrix} = \frac{1}{\Delta}\begin{bmatrix} K_2 P_1 - R_3 P_2 \\ -R_3 P_1 + c_{44} P_2 \end{bmatrix}$$

we get from (27)

$$\mathbf{u}_{,1}(x_1,0) = \frac{1}{\pi\Delta\sqrt{x_1^2 - b^2}}\begin{bmatrix} K_2 P_1 - R_3 P_2 \\ -R_3 P_1 + c_{44} P_2 \end{bmatrix} \text{ for } x_1 > b, \quad (30)$$

where $\Delta = c_{44}K_2 - R_3^2$. Thus, due to (9) and (30) one has:

$$\frac{\partial u_3}{\partial x_1}(x_1,0) = \frac{K_2 P_1 - R_3 P_2}{\pi\Delta\sqrt{x_1^2 - b^2}}, \ \frac{\partial w_3}{\partial x_1}(x_1,0) = \frac{R_3 P_1 - c_{44} P_2}{\pi\Delta\sqrt{x_1^2 - b^2}} \text{ for } x > b \quad (31)$$

Find next the phonon and phason displacements for $/x_1/ > b$ by integrating the formulas (31). We take into account that

$$I(x_1) = \int\frac{dx_1}{\sqrt{x_1^2 - b^2}} = \log\left(|x_1| + \sqrt{x_1^2 - b^2}\right) + \log c^*.$$

Defining the arbitrary constant $c^*$ from the condition $I(\pm b) = 0$, we get:

$$I(x_1) = \log\left(\frac{|x_1|}{b} + \sqrt{\frac{x_1^2}{b^2} - 1}\right). \quad (32)$$

Substituting the last formula into the formulas (31) leads to

$$\mathbf{u}(x_1,0) = \frac{1}{\pi\Delta}\begin{bmatrix} K_2 P_1 - R_3 P_2 \\ R_3 P_1 - c_{44} P_2 \end{bmatrix}\log\left(\frac{|x_1|}{b} + \sqrt{\frac{x_1^2}{b^2} - 1}\right) \text{ for } x > b. \quad (33)$$

It follows from the formula (32) that $\mathbf{u}(b,0) = 0$.

Analysis of the function (32) shows that for $x_1 \to b + 0$ one has:

$$I(x_1)\big|_{x_1\to b+0} \to \log\left(1+\sqrt{\frac{x_1^2}{b^2}-1}\right) \to \sqrt{\frac{x_1^2}{b^2}-1} \to \sqrt{2}\sqrt{\frac{|x_1|}{b}-1}.$$

It means that $u(x_1,0)$ has the following behavior for $x_1 \to b+0$:

$$u(x_1,0) \to \frac{\sqrt{2}}{\pi\Delta}\begin{bmatrix} K_2P_1-R_3P_2 \\ R_3P_1-c_{44}P_2 \end{bmatrix}\sqrt{\frac{|x_1|}{b}-1} \qquad (34)$$

**Numerical analysis and discussion.** Consider the quasicrystalline material with the following characteristics [2]:

$K_2 = 0.15\times 10^9$ Pa, $R_3 = 1.76\times 10^9$ Pa, $C_{44} = 3.55\times 10^{10}$ Pa

and assume that b=0.01 m. Next we illustrate different phonon and phason values obtained for different magnitudes of phonon and phason loadings.

In Fig. 3 the variations of phonon (rigid lines) and phason (dashed lines) stresses along the lover edge of the beam are presented. The lines I were obtained for $\mathbf{P}^{(1)} = \begin{Bmatrix} P_1^{(1)} \\ P_2^{(1)} \end{Bmatrix} = \begin{Bmatrix} 10^4 \\ 4\times 10^3 \end{Bmatrix} N/m$ (first case of the loading) and the lines II were obtained for $\mathbf{P}^{(2)} = \begin{Bmatrix} P_1^{(2)} \\ P_2^{(2)} \end{Bmatrix} = \begin{Bmatrix} 6\times 10^3 \\ 2\times 10^3 \end{Bmatrix} N/m$ (second case of the loading).

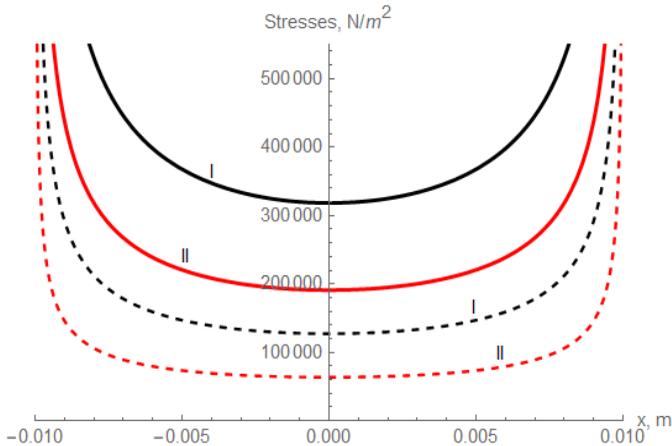

Fig. 3. **Variations of phonon (rigid lines) and phason (dashed lines) stresses along the lover edge of the beam for two cases of the external loadings**

In Fig. 4 the variations of phonon (rigid lines) and phason (dashed lines) derivatives of displacements along the edge of the half-plane outside of the beam are given for the same loadings as in Fig. 3. The lines I and II correspond to these cases of the loadings. For graphic clarity the results correspondent to the phason lines are scaled down by a factor 5.

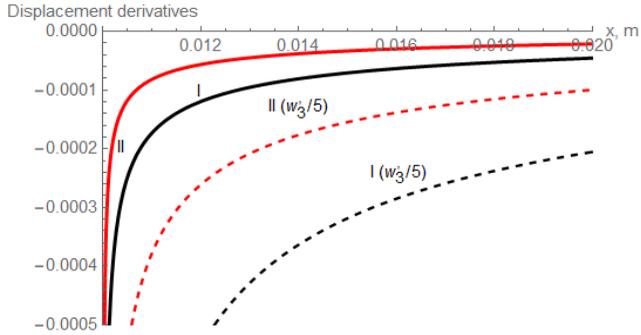

Fig. 4. **The variations of phonon (rigid lines) and phason (dashed lines) derivatives of displacements along the edge of the half-plane outside of the beam**

In Fig. 5, the displacements of both the phonon (solid lines) and phason (dashed lines) fields are shown along the edge of the half-plane outside of the beam for the same cases of external loading as in Fig. 3. The lines I, correspond to the first loading case, while the lines II, correspond to the second loading case. The phonon displacements represent the lattice vibrations within the material, while the phason displacements indicate the quasi-periodic rearrangement of atoms in the quasicrystal structure. The phonon displacements change more smoothly and evenly along the considered interval, while the phason displacements decrease more sharply, particularly close to the beam. The phonon and phason displacements differ significantly in magnitude due to the distinct elastic properties of the fields, with phason displacements being significantly larger. For this reason, the phason displacements are scaled down by a factor of 5 in the graph to facilitate better visualization.

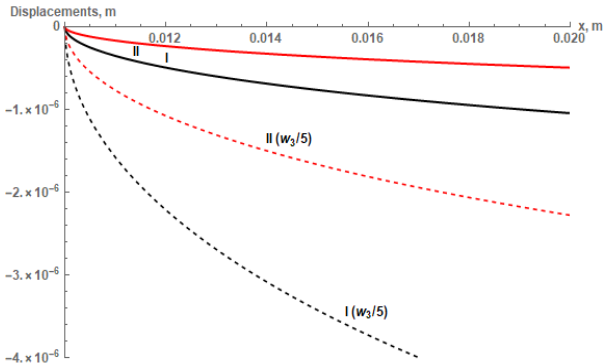

Fig. 5. **The variations of phonon (rigid lines) and phason (dashed lines) displacements along the edge of the half-plane outside of the beam**

**Conclusion.** The anti-plane problem of the contact of an absolutely rigid beam and a quasi-crystalline half-space is considered. Representations of the phonon and phason components of the stress-strain state through a piecewise analytical vector-function are developed. Based on these representations, the problem is reduced to a vector problem of linear relationship, which solution is given in analytical form. It was established that the stresses and derivatives of displacements have root singularities when approaching the corner points of the beam. For a specific quasi-crystalline material and various loads, the graphs of stresses, displacements and their derivatives variations along the corresponding parts of the half-plane boundary are constructed. It was found that for the same load, phonon stresses are, as a rule, greater than phason stresses. As for displacements, their phason components are several times larger than the corresponding phonon factors. It should also be noted that there is a sharp increase in stresses and derivatives of displacements when approaching the corner points of the beam.